\documentclass{llncs} 

\usepackage{
        ,latexsym
        ,amsmath
        ,amssymb
        ,amsfonts
	,graphicx
        }

\title{Effective Privacy Amplification for Secure Classical Communications\thanks{To appear in {\it Europhysics Letters.}}}

\author{
	Tam\'as Horv\'ath\inst{1,2} \and
	Laszlo B. Kish\inst{3} \and
	Jacob Scheuer~\inst{4,5}
}		

\institute{                    
Department of Computer Science, University of Bonn, Germany
\and
Fraunhofer IAIS, Schloss Birlinghoven, D-53754 Sankt Augustin, Germany 
\email{tamas.horvath@iais.fraunhofer.de} \medskip
\and
Department of Electrical and Computer Engineering, Texas A\&M University, College Station, TX 77843-3128, USA \\
\email{laszlo.kish@mail.ece.tamu.edu} \medskip
\and 
School of Electrical Engineering, Tel-Aviv University, Ramat-Aviv, Israel 
\and
Center for Nanoscience and Nanotechnology, Tel-Aviv University, \\ Ramat-Aviv, Israel \\
\email{kobys@eng.tau.ac.il}
}

\begin{document}
\frontmatter
\pagestyle{headings}
\mainmatter
\maketitle

\abstract{
We study the practical effectiveness of privacy amplification for classical key-distribution
schemes. We find that in contrast to quantum key distribution schemes, the high fidelity
of the raw key generated in classical systems allow the users to always sift a secure
shorter key if they have an upper bound on the eavesdropper probability to correctly guess
the exchanged key-bits. The number of privacy amplification iterations
needed to achieve information leak of $10^{-8}$ in existing classical communicators is 2 or 3 resulting in a corresponding slowdown $4$ to $8$.
We analyze the inherent tradeoff between the number of iterations and the security of the raw key. This
property which is unique to classical key distribution systems render them highly useful
for practical, especially for noisy channels where sufficiently low quantum bit error ratios
are difficult to achieve.
}

\section{Introduction}

The ability of a secure communication system to protect the information being transferred
through it is primarily determined by the encryption key and the ability of the users
(Alice and Bob) to keep it secret from a potential eavesdropper (Eve). In fact, under the
strictest conditions, the key is the only element of the encryption scheme which is
assumed to be unknown to the adversaries and, as such, constitute the last and only line
of defense of the data. Although absolute information security requires an encryption key
which is as long as the message (i.e., the one-time-pad)~\cite{b15}, it is clear that for practical
reasons it is desired to employ a shorter key. Regardless the specific encryption
algorithm, achieving higher security level requires that the adversary has minimal
information on the encryption key.

The most profound challenge in achieving secure communication is probably the
distribution of this key between the users~\cite{b15}. Secure communication between the users
cannot take place without a key, but the distribution of this key seems to necessitate the
existence of a secure communication channel. In attempt to resolve this loophole, much
effort has been focused on the development of physical-layer key-distribution scheme.
Quantum key distribution (QKD)~\cite{b16,b17} is probably the most well known scheme although
recently, alternative concepts employing classical physics have been proposed and
demonstrated~\cite{b1,b2,b3,b4,b5,b6,b7,b8,b9,b10,b11}. 
For example, the demonstration of the Kirchhoff-Law-Johnson-(like)-Noise based key distribution system at 2000km range resulted in $99.98\%$ fidelity~\cite{b3}, or recent studies on ultralong fiber laser classical key distribution
schemes have demonstrated ranges exceeding 200km with potential key establishing
rates of hundreds bits per second~\cite{b18}.

\begin{table}[t]
\caption{The physical quantities characterizing the channel state in the three classical key
distribution schemes and the distinguishable states.}
\label{tab1}
\begin{center}
\begin{tabular}{ll}
{\bf Method} & {\bf Channel signal (the physical quantity of the channel state)} \\
 & {\bf and the types of states}  \\ \hline
KLJN & mean-square voltage noise or current noise amplitude, \\
 & $HH$, $LL$, $LH=HL=S$ \\[0.5em] 
UFL & frequency and power of lasing, \\
&  $HH$, $LL$, $LH=HL=S$ \\[0.5em]
Liu & cross-correlation of noise signals emitted by Alice and Bob,  \\
&  $HH=LL$, $LH=HL=S$
\end{tabular}
\end{center}
\end{table}

Whether quantum or classical, all the physical-layer key-distribution schemes provide a
relatively low key establishing rate. Therefore, for practical reasons, Alice and Bob
cannot afford discarding a partially exposed key and it is, thus, desired to develop an
algorithm to purify a partially exposed key even at the expense of the key length.

Privacy amplification (PA)~\cite{b19} is an example for an algorithm which provides this ability.
One of the simplest PA algorithms is replacing the original key by a shorter one which
bits are the product of the XOR operation between two successive bits of the original key (see, e.g., \cite{b20}). 
The length of the new key is half of that of the original one but Eve's knowledge on the
new key is substantially reduced (see section 3). 
We mathematically analyze this simple algorithm and show for the three existing classical key-distribution
schemes~\cite{b1,b2,b3,b4,b5,b6,b7,b8,b9,b10,b11} that a practically satisfactory approach of $0.5$ 
(i.e., zero information) can be reached by $4$--$8$ times slowdown for the state-of-the-art physical realizations of these systems.
Notice that for eavesdropper probability $0.5$, Eve obtains indeed zero information on the key because this probability is equal to generating her own key bits by using an unbiased random coin.

Although PA is essentially a classical
scheme, it was developed primarily as an accessory tool for QKD in order to purify a
partially exposed key. However, because of the fragile nature of quantum states the very
same one which provides Alice and Bob with the ability to detect the presence of Eve, the
ability of Alice and Bob to purify their key is limited by the quantum bit error ratio
(QBER). In particular, it has been shown that if the QBER exceeds a certain value (which
depends on the particular QKD scheme), then classical PA cannot be used to sift a
shorter, secure key from the raw one. 

On the other hand, in classical key distribution schemes Alice and Bob are unable to
detect a passive eavesdropper because, unlike QKD, the activity of such adversary does
not necessarily induce errors in the key-exchange protocol.

However, Alice and Bob can set an upper bound $p$ on the probability of Eve to correctly
identify an exchanged key-bit and therefore, can set a bound on the maximal knowledge
Eve could have gained on the raw key. This bound can be in turn used in a PA scheme to
reduce Eve knowledge on the key to any desired level provided that the fidelity $F$ of the
key exchanged between Alice and Bob is sufficiently large to allow that.

In this paper, we study the employment of PA to classical key-distribution schemes. We
show that the inherent robustness of such scheme -- the one which prevents Alice and
Bob from detecting the presence of Eve, provides them with the ability to sift a secret key
from the raw one for any exposure probability as long as the fidelity $F = 1$. The direct
implication is that classical key distribution schemes can facilitate secure communication
in difficult and noisy channel conditions which are beyond the capabilities of QKD
schemes. In section 2 we briefly describe the properties of classical key-distribution
schemes. In section 3 we describe and analyze the PA scheme we use and in section 4 we
summarize the results and conclude.

\section{Secure classical key exchange protocols}

Currently there are three classical secure key distribution methods which have been published; in chronological order: the Kirchhoff-Law-Johnson-(like)-Noise (KLJN)~\cite{b1,b2,b3,b4}, the Ultra-long-Fiber-Laser (UFL)~\cite{b5,b6,b7,b8}, and the Pao-Lo Liu (Liu)~\cite{b9,b10,b11} schemes. In all these schemes, Alice and Bob are interacting with the channel via a classical physical quantity while they are monitoring some properties of the channel (i.e. the channel ``status''). The channel status depends on the actions of Alice and Bob. Alice and Bob can choose between two different types of operations on the channel, which correspond to the $H$ and the $L$ bits. Choosing different interaction types, namely $HL$ or $LH$, results in identical channel state $S$ while the other action combinations, namely $HH$ and $LL$ result in different channel states. Eve cannot directly observe Alice's and Bob's actions on the channel but she can learn the interaction types in the case of $LL$ and $HH$. In the $LH$ and $HL$ cases, Eve does not know which side has the $L$ and which one the $H$. However, Alice and Bob know their own bit value and thus, when they observe the $S$ channel state, they know that the other side has the opposite bit. 
It is important to note that, in these systems, the fidelity of key exchange between Alice and Bob can approach 100\% depending on how good statistics (how long clock time) are devoted to the exchange of a single bit. 
The exception from this situations is the case of invasive attack by Eve~\cite{b15,b16} which will be addressed at the end of this section.

Table~\ref{tab1} shows the physical quantities characterizing the channel state in the three classical key distribution schemes mentioned above. The KLJN and UFL systems have 3 different distinguishable states while the Liu protocol has 2.

It should be noted that the successful operation of the three classical scheme requires a high degree of temporal synchronization between the actions taken by Alice and Bob.
Though this is doable with today's technology, practical solutions, with ramping the signals up \cite{b3} at the beginning of the clock signal and down at the end of that, do not require perfect synchronization.

\begin{figure}[t]
\begin{center}
\includegraphics[scale = 0.15]{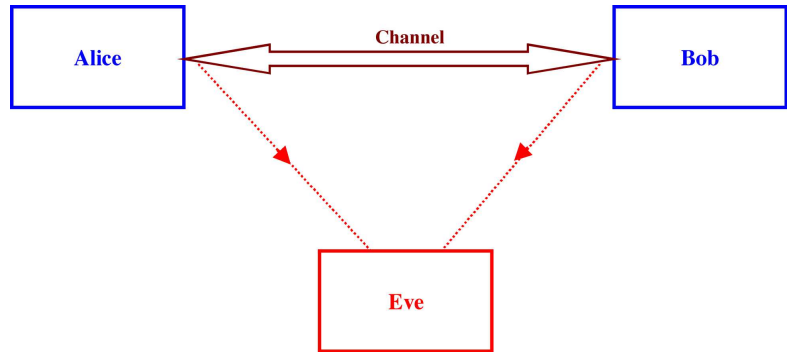}
\end{center}
\caption{To utilize the small information leak due to non-idealities, Eve measures and compares the
channel signals at the two ends (KLJN) or the propagating signals in the two directions (UFL and Liu).}
\label{fig1}
\end{figure}

In a practical physical secure layer, due to non-idealities, finite clock speed, etc., there is some information leak~\cite{b21}, which Eve can utilize to gain information, see Figure~\ref{fig1}. In the KLJN system~\cite{b1,b2,b3,b4} Eve can utilize the small difference of the noise signal at the two ends of the wire \cite{b12,b13} due to wire resistance, capacitance and inductance. In the UFL scheme~\cite{b5,b6,b7,b8}, comparison of the spectra serves as information for Eve~\cite{b5}. Similarly, in the Liu scheme~\cite{b9,b10,b11}, there is some information about the bits of Alice and Bob in the interrelation of the signals sent/reflected by Alice and Bob, respectively~\cite{b9,b10,b11,b14}.

\begin{figure}[t]
\begin{center}
\includegraphics[scale = 0.1]{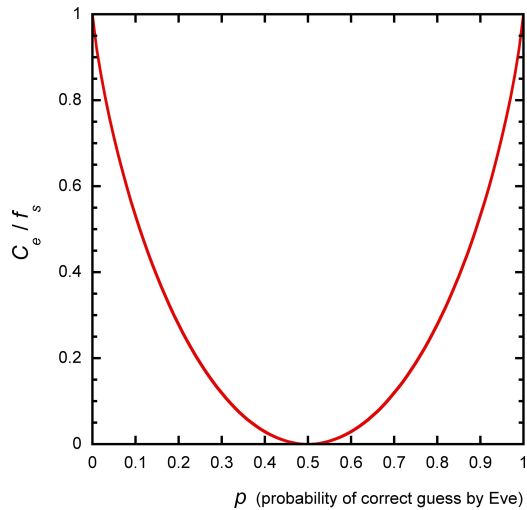}
\end{center}
\caption{$C_e/f_s$ is the fraction of information leaked by each exchanged key bit toward Eve versus the probability of her correct guess. In the case of error-free key exchange between Alice and Bob, $C_e/f_s$ is the ratio of information channel capacity between Alice/Bob toward Eve and that of the key exchange between Alice and Bob.}
\label{fig2}
\end{figure}

In the ideal case, where there is no information leak, Eve's success probability of guessing the key bits is $p=0.5$ which means that she obtains zero information on the key. In realistic cases, the actual information leak toward Eve~\cite{b3} can be calculated by the Shannon formula for digital channels~\cite{b21}:
$$
C_e = f_s \left[ 1+ p \log_2{p} + (1-p) \log_2{(1-p)}\right] ,
$$
where $C_e$ is the information channel capacity,
$f_s$ is the frequency of exchanged key bits,
$p$ is the probability of correct bit guess by Eve and
$1- p$ is her error probability (see, also, Figure~\ref{fig2}).

As a practical goal of PA in the present paper, we set the limit of $$C_e / f_s \leq 10^{-8} \enspace ,$$ which corresponds to 
\begin{equation}
\label{eq:p}
p \approx 0.5006 \enspace.
\end{equation}

\begin{figure}[t]
\begin{center}
\includegraphics[scale = 0.15]{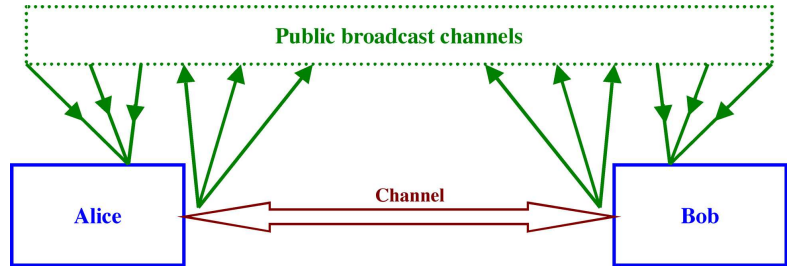}
\end{center}
\caption{Alice and Bob are broadcasting and comparing the channel signals (analog quantities) seen by them to defend against any invasive attack and to approach 100\% fidelity.}
\label{fig3}
\end{figure}

Another important aspect of security of these secure key exchange protocols is the defense against invasive attacks~\cite{b1} including the man-in-the-middle attack~\cite{b2}. In order to have an effective protection of each of the key bits, they broadcast the channel signal~\cite{b1,b2} in as many public channels as possible and compare the broadcasted signals, see Figure~\ref{fig3}. As a result they have complete information about the channel signal seen at the other end. During the comparison they can detect any invasive manipulation on the channel signal that can be utilized by Eve and thus they can discard the related bit if necessary~\cite{b1,b2}.

For PA, the topic of this paper, this broadcasting is highly beneficial because it allows Alice and Bob to know exactly what would be the decision of the other end on the channel state: secure or not. If they see that their decision is going to be different, they discard the bit. Thus, in this way they rectify the errors in the key exchange between them and approach 100\% fidelity.

\section{The PA algorithm}
In this section we first describe the privacy amplification algorithm based on iteratively replacing the original key by shorter one using the binary XOR operation and then demonstrate its practical effectiveness for the state-of-the-art  secure classical key exchange schemes such as the KLJN, Liu, and UFL. 
The input to the algorithm consists of the following three parameters:
\begin{itemize}
\item[(i)]
the length $L > 0$ of the key to be generated,
\item[(ii)]
an upper bound $p$ ($0.5 \leq p < 1$) on the probability of correct identification of a bit by Eve, and
\item[(iii)]
a user defined upper bound $\epsilon > 0$ on the distance from the zero information point $0.5$.
\end{itemize}
We assume that the values for these input parameters are known for all parties of the communications, including Eve. {\em Given} $L$, $p$, and $\epsilon$, the {\em goal} of Alice and Bob is to generate and exchange a secure key of length $L$ such that the probability of Eve to correctly identify any particular bit of the key is at most $0.5 + \epsilon$. 

Our estimations of $p$ (i.e., Eve's probability to correctly guess the bits of the raw key) are based on experimental and simulation studies for the three key distribution systems:
\begin{itemize}
\item 
For the KLJN scheme $p$ was experimentally studied in a 2000km laboratory model line~\cite{b3}.
It was found to be $0.525$ with fidelity of $99.98\%$. 
\item 
For the UFL scheme a full-scale $50$km system was studied experimentally~\cite{b7}, yielding a raw-key guess probability of $0.65$ with $99.4\%$ fidelity.  
\item
For the Liu scheme, the raw-key guess probability was studied by performing computer simulations in MATLAB for various values of the system parameters~\cite{b9}. For the strategically identified parameter values reported in \cite{b9}, $p$ was found to be $0.573$ with $91.8\%$ fidelity.
\end{itemize}
For $\epsilon$, we use the value $0.0006$  based on the practical considerations discussed in the previous section (see, also,  (\ref{eq:p})).

Given $p$ and $\epsilon$, Alice and Bob first calculate a positive integer $k = k(p,\epsilon)$ defined below and then generate, for each bit $B_i$ of the final key ($1 \leq i \leq L$), a sequence $S_i$ consisting of $2^k$ raw bits. By raw bits we mean bits with eavesdropping probability at most $p$. High risk bits, i.e., for which this probability is larger than $p$, are disregarded (see, e.g., KLJN~\cite{b1}).  

 From $S_i^0=S_i$, Alice and Bob compute the binary strings
\begin{eqnarray*}
S_i^1 & = &  F(S_i^0) \\ 
S_i^2 &= & F(S_i^1) \\ 
& \vdots \\
S_i^k & = & F(S_i^{k-1})
\end{eqnarray*}
and define the $i$th bit $B_i$ of the final key to be $S_i^k$, where 
$$
F(b_1 b_2 b_3 \ldots b_{2n}) = b_1' b_2' \cdots b_n' \textrm{ with } b_j' = \textrm{XOR}(b_{2j-1}, b_{2j})
$$ 
for every $j = 1,\ldots,n$.

We first show that the probability of Eve to correctly guess the key-bits decreases monotonically with each iteration of the PA algorithm. More precisely, we prove by induction on $l$ that, for every bit $B$ of $S_i^l$,
\begin{equation}
\label{eq:ub}
\textrm{\bf Pr}(\textrm{Eve correctly identifies $B$}) \leq P^l(p)
\end{equation}
 for all $l \geq 1$, where 
$$
P(x) = 2x^2-2x+1
$$
and
$$
P^l(x) =
\begin{cases}
P(x) & \text{if $l = 1$} \\
P(P^{l-1}(x)) & \text{o/w.}
\end{cases}
$$
For the base case $l=1$, any bit $b'_j = \textrm{XOR}(b_{2j-1},b_{2j})$ of $S^1_i$ is correctly identified  either by correctly identifying both raw bits $b_{2j-1}$ and $b_{2j}$ or by incorrectly identifying both of them, as $\textrm{XOR}(b_{2j-1},b_{2j}) = \textrm{XOR}(\overline{b}_{2j-1},\overline{b}_{2j})$.
Thus, assuming that Eve correctly identifies a raw bit with probability $q$, the probability that Eve correctly identifies $b'_j$ is 
$$ q^2+(1-q)^2 = P(q) \enspace .
$$
Since $0.5 \leq q \leq p$, we have $P(q) \leq P(p)$ and thus, (\ref{eq:ub}) holds for $l=1$. 
The proof for the induction step can be shown by using similar arguments.

We now show that $\lim\limits_{l \to \infty} P^l(p) = 0.5$ for every $p \in [0.5,1)$.
One can easily check that
\begin{itemize}
\item[(i)] $P(x) \leq x$ holds for every $x \in [0.5,1)$ and
\item[(ii)] $P(x)$ is strictly monotonically increasing on the interval $[0.5,1)$.
\end{itemize}
Thus, for every $p \in [0.5,1)$, the sequence 
\begin{equation}
\label{eq:seq}
p, P(p), P^2(p), P^3(p),\ldots
\end{equation}
is strictly monotonically decreasing. To show that it converges to the desired value $0.5$,
let 
$$\Delta(x) = x - P(x) = -2x^2+3x -1$$
and consider the case that $0.75 \leq p <1$.
Since $\Delta$ is (strictly) monotonically decreasing on the interval $[0.75,1)$, there must be a positive integer  
$$l \leq \left \lceil \frac{p-0.75}{\Delta(p)}\right \rceil$$ satisfying $P^l(p) < 0.75$. 

Let $l_0$ be the smallest such integer if $p > 0.75$; otherwise let $l_0 = 0$.
To show that the sequence in (\ref{eq:seq}) is convergent, it suffices to show that the subsequence
\begin{equation}
\label{eq:subseq}
P^{l_0}(p), P^{l_0+1}(p),\ldots
\end{equation}
is convergent.

We first note that, for every $0.5 \leq x_1,x_2 < 0.75$ we have
\begin{eqnarray}
|P(x_1)-P(x_2)| 
& = & 2|(x_1 - x_2)(x_1+x_2-1)| \notag \\
& \leq & q|x_1 - x_2| \label{eq:lip}
\end{eqnarray}
for some positive real number $q < 1$. Thus, $P$ is a {\em contraction} function on the interval $[0.5,P^{l_0}(p)]$ and hence, by the Banach fixpoint theorem\footnote{
Let $f : [a,b] \to [a,b]$ be a contraction function, i.e., there exists a positive real number $q < 1$ (Lipschitz constant) such that $|f(x)-f(y)| \leq q|x-y|$ for every $a \leq x, y \leq b$. 
Then $f$ has exactly one fixpoint $x^* \in [a,b]$ and $x^* = \lim\limits_{n \to \infty} f^n(s)$ for any $s \in [a,b]$. 
}, the sequence in (\ref{eq:subseq}) converges to the fixpoint $0.5$ of $P$. Putting all these together, we have that
$$
\lim_{n \to \infty} P^n(p) = 0.5
$$
for every $p \in [0.5,1)$, i.e., for sufficiently large $n$, the probability that Eve correctly identifies any bit of the final key is close to $0.5$.  

We now turn to the choice of the number $k(p,\epsilon)$ satisfying 
$$
P^{k(p,\epsilon)}(p) \leq 0.5 + \epsilon
\enspace .
$$
For the elements of the sequence in (\ref{eq:subseq}) we have 
$$
P^{l_0+l}(p) -0.5 \leq  \frac{q^l}{1-q}\left(P^{l_0}(p) -P^{l_0+1}(p)\right)  \leq \epsilon
$$
for sufficiently large $l \geq 0$, where for the  Lipschitz constant $q$ we have 
$$
q = 4P^{l_0}(p) -2
$$
by using (\ref{eq:lip}).
Notice that $q < 1$, as $P^{l_0}(p) < 0.75$ by the choice of $l_0$. Thus, for the convergence rate of the sequence in (\ref{eq:seq}) we get the upper bound
$$
k(p,\epsilon) \leq \left\lceil \max\left(0,\frac{p-0.75}{\Delta(p)}\right) +  \log_q \frac{(1-q)\epsilon}{\Delta(P^{l_0}(p))}\right\rceil 
\enspace .
$$
Given $p$ and $\epsilon$, one can easily find the exact value of $k(p,\epsilon)$. For the particular value of $\epsilon = 0.0006$ discussed above, we give $k(p,\epsilon)$ for different $p$'s  in Table~\ref{tab2}, including $p = 0.525$ (KLJN~\cite{b3}), $p=0.573$ (Liu~\cite{b9}), and $p=0.65$ (UFL~\cite{b8}).

\begin{table}[t]
\begin{center}
\caption{The values for $k= k(p,0.0006)$ and $P^{k}(p)$ (rounded to five places) for different values of $p$, including those for the particular realizations of  KLJN, Liu, and UFL. }
\label{tab2}
\begin{tabular}{lcl}
\hline
\multicolumn{1}{c}{$p$} & $k=k(p,0.0006)$ & $P^k(p)$ \\ \hline
0.99 & 9 & 0.50002 \\
0.90 & 6 & 0.50040 \\
0.85 & 5 & 0.50001 \\
0.80 & 4 & 0.50014 \\
0.70 & 4 & 0.50000 \\
\textbf{0.65 (UFL)} & \textbf{3} & 0.50003 \\
0.60 & 3 & 0.50000 \\
\textbf{0.573 (Liu)} & \textbf{2} & 0.50023 \\
0.55 & 2 & 0.50005 \\
\textbf{0.525 (KLJN)} & \textbf{2} & 0.50000
\end{tabular}
\end{center}
\end{table}

As expected, $k$ decreases with $p$. The practical significance of our privacy amplification algorithm is clearly demonstrated for the particular realizations of the classical key-distribution schemes. For these physical realizations, the generation of a secure bit requires only 4 raw bits for the KLJN  cipher and the Liu  and 8 for the UFL.

\section{Conclusions}

We have studied the effectiveness of a privacy amplification algorithm for classical key-distribution schemes. We have found that because of the high fidelity of the raw key which is available to Alice and Bob, they can always sift a secure, shorter key as long as they can find an upper bound of Eve's probability to correctly guess the exchanged key-bits. This is in contrast to QKD schemes where the PA is limited to situations where the QBERs are lower than a certain value ($\sim 27\%$ for the best QKD scheme). This profound difference stems from Alice's and Bob's ability to achieve complete fidelity of their key when using a classical key-distribution scheme. Unlike QKD systems, such complete fidelity can be achieved if Alice and Bob publish information on their measurements, thus allowing each party to deduce the conclusion of the other one (secure or non-secure bit exchange). In QKD schemes, detector dark-counts, background noise and Eve's interference with the channel necessarily generates errors, which reduce the fidelity of the raw key.
Regardless Eve's success probability, any desired level of information leak can be achieved by sufficient number of iterations of the PA algorithm, where there is a clear tradeoff between the ``raw'' security of the system (manifested by Eve's success probability, $p$) and the required number of iterations. For example, using the experimentally estimated leakage of the key-bits towards Eve for the UFL, the Liu, and the KLJN schemes, we found that information leak level of $10^{-8}$ can be achieved by successively applying the PA three times for the first and twice for the second and the third.
The ability of Alice and Bob to always sift a secure key is an important strength of the classical key-distribution schemes, which render them highly attractive for practical applications, in particular, for noisy channels where sufficiently low QBER is difficult to achieve.

\subsection*{Acknowledgments}
Research carried out at Texas A\&M University during the short visit of T. Horv\'ath and J. Scheuer.
The visit and stay was partially supported by the Texas A\&M subcontract from Signal Processing Inc. on Fluctuation-Enhanced Sensing.

\bibliographystyle{abbrv}

\end{document}